\documentstyle[12pt]{article}
\input epsf

\begin{document}
\baselineskip=20.5pt

\def\beqra{\begin{eqnarray}} \def\eeqra{\end{eqnarray}}
\def\beqast{\begin{eqnarray*}} \def\eeqast{\end{eqnarray*}}
\def\beq{\begin{equation}}      \def\eeq{\end{equation}}
\def\be{\begin{enumerate}}   \def\ee{\end{enumerate}}

\def\fnote#1#2{\begingroup\def\thefootnote{#1}\footnote{#2}\addtocounter
{footnote}{-1}\endgroup}

\def\ut#1#2{\hfill{UTTG-{#1}-{#2}}}
\def\fl#1#2{\hfill{FERMILAB-PUB-94/{#1}-{#2}}}
\def\itp#1#2{\hfill{NSF-ITP-{#1}-{#2}}}

\def\bet{\beta}
\def\gam{\gamma}
\def\Gam{\Gamma}
\def\la{\lambda}
\def\eps{\epsilon}
\def\La{\Lambda}
\def\si{\sigma}
\def\Si{\Sigma}
\def\al{\alpha}
\def\Tha{\Theta}
\def\tha{\theta}
\def\vphi{\varphi}
\def\del{\delta}
\def\Del{\Delta}
\def\ab{\alpha\beta}
\def\om{\omega}
\def\Om{\Omega}
\def\mn{\mu\nu}
\def\mun{^{\mu}{}_{\nu}}
\def\kap{\kappa}
\def\rsi{\rho\sigma}
\def\beal{\beta\alpha}

\def\til{\tilde}
\def\rta{\rightarrow}
\def\eqv{\equiv}
\def\nab{\nabla}
\def\pa{\partial}
\def\sit{\tilde\sigma}
\def\ul{\underline}
\def\indt{\parindent2.5em}
\def\nd{\noindent}

\def\rsi{\rho\sigma}
\def\beal{\beta\alpha}

\def\caa{{\cal A}}
\def\cb{{\cal B}}
\def\cac{{\cal C}}
\def\cd{{\cal D}}
\def\ce{{\cal E}}
\def\cf{{\cal F}}
\def\cg{{\cal G}}
\def\cah{{\cal H}}
\def\ci{{\cal I}}
\def\cj{{\cal{J}}}
\def\ck{{\cal K}}
\def\cl{{\cal L}}
\def\cm{{\cal M}}
\def\cn{{\cal N}}
\def\cO{{\cal O}}
\def\cp{{\cal P}}
\def\car{{\cal R}}
\def\cs{{\cal S}}
\def\ct{{\cal{T}}}
\def\cu{{\cal{U}}}
\def\cv{{\cal{V}}}
\def\cw{{\cal{W}}}
\def\cx{{\cal{X}}}
\def\cy{{\cal{Y}}}
\def\cz{{\cal{Z}}}

\def\cdgc{C^{\dagger}C}
\def\ccdg{CC^{\dagger}}
\def\pcdgc{C'^{\dagger}C'}
\def\pccdg{C'C'^{\dagger}}
\def\vdgv{v^{\dagger}v}

\def\raisenot{\raise .5mm\hbox{/}}
\def\nota{\ \hbox{{$a$}\kern-.49em\hbox{/}}}
\def\notA{\hbox{{$A$}\kern-.54em\hbox{\raisenot}}}
\def\notb{\ \hbox{{$b$}\kern-.47em\hbox{/}}}
\def\notB{\ \hbox{{$B$}\kern-.60em\hbox{\raisenot}}}
\def\notc{\ \hbox{{$c$}\kern-.45em\hbox{/}}}
\def\notd{\ \hbox{{$d$}\kern-.53em\hbox{/}}}
\def\notbd{\ \hbox{{$D$}\kern-.61em\hbox{\raisenot}}} 
\def\note{\ \hbox{{$e$}\kern-.47em\hbox{/}}}
\def\notk{\ \hbox{{$k$}\kern-.51em\hbox{/}}}
\def\notp{\ \hbox{{$p$}\kern-.43em\hbox{/}}}
\def\notq{\ \hbox{{$q$}\kern-.47em\hbox{/}}}
\def\notW{\ \hbox{{$W$}\kern-.75em\hbox{\raisenot}}}
\def\notz{\ \hbox{{$Z$}\kern-.61em\hbox{\raisenot}}}
\def\notpa{\hbox{{$\partial$}\kern-.54em\hbox{\raisenot}}}

\def\fo{\hbox{{1}\kern-.25em\hbox{l}}}  
\def\rf#1{$^{#1}$}
\def\bx{\Box}
\def\tr{{\rm Tr}}
\def\trm{{\rm Tr}_{_{\left(M\right)}}~}
\def\trn{{\rm Tr}_{_{\left(N\right)}}~}
\def\trnn{{\rm Tr}_{_{\left(N+1\right)}}~}
\def\rmtr{{\rm tr}}
\def\dgg{\dagger}

\def\lag{\langle}
\def\rag{\rangle}
\def\bmid{\big|}

\def\vlap{\overrightarrow{\La p}} 
\def\lrta{\longrightarrow} \def\lrar{\raisebox{.8ex}{$\longrightarrow$}}
\def\rlarw{\longleftarrow\!\!\!\!\!\!\!\!\!\!\!\lrar}

\def\llra{\relbar\joinrel\longrightarrow}           
\def\mapright#1{\smash{\mathop{\llra}\limits_{#1}}} 
\def\mapup#1{\smash{\mathop{\llra}\limits^{#1}}} 

\def\nmasymptotic{
{_{\displaystyle{\rm lim}}\atop
{\scriptstyle N,M\rightarrow\infty}
}\,\, 
}

\def\nasymptotic{{_{\stackrel{\displaystyle\longrightarrow}
{N\rightarrow\infty}}\,\, }} 
\def\masymptotic{{_{\stackrel{\displaystyle\longrightarrow}
{M\rightarrow\infty}}\,\, }} 
\def\wasymptotic{{_{\stackrel{\displaystyle\longrightarrow}
{w\rightarrow\infty}}\,\, }} 

\def\asymptext{\raisebox{.6ex}{${_{\stackrel{\displaystyle\longrightarro
w}{x\rightarrow\pm\infty}}\,\, }$}} 

\def\7#1#2{\mathop{\null#2}\limits^{#1}}        
\def\5#1#2{\mathop{\null#2}\limits_{#1}}        
\def\too#1{\stackrel{#1}{\to}}
\def\tooo#1{\stackrel{#1}{\longleftarrow}}
\def\nout{{\rm in \atop out}}

\def\one{\raisebox{.5ex}{1}}
\def\BM#1{\mbox{\boldmath{$#1$}}}

\def\ltsim{\matrix{<\cr\noalign{\vskip-7pt}\sim\cr}}
\def\gtsim{\matrix{>\cr\noalign{\vskip-7pt}\sim\cr}}
\def\haf{\frac{1}{2}}


\def\place#1#2#3{\vbox to0pt{\kern-\parskip\kern-7pt
                             \kern-#2truein\hbox{\kern#1truein #3}
                             \vss}\nointerlineskip}

\def\illustration #1 by #2 (#3){\vbox to #2{\hrule width #1 height 0pt 
depth
0pt
                                       \vfill\special{illustration #3}}}

\def\scaledillustration #1 by #2 (#3 scaled #4){{\dimen0=#1 \dimen1=#2
           \divide\dimen0 by 1000 \multiply\dimen0 by #4
            \divide\dimen1 by 1000 \multiply\dimen1 by #4
            \illustration \dimen0 by \dimen1 (#3 scaled #4)}}

\def\ON{{\cal O}(N)}
\def\UN{{\cal U}(N)}
\def\bdPh{\mbox{\boldmath{$\dot{\!\Phi}$}}}
\def\bPh{\mbox{\boldmath{$\Phi$}}}
\def\bPhs{\bPh^2}
\def\sef{S_{eff}[\sigma,\pi]}
\def\sigx{\sigma(x)}
\def\pix{\pi(x)}
\def\bph{\mbox{\boldmath{$\phi$}}}
\def\bphs{\bph^2}
\def\ex{\BM{x}}
\def\exs{\ex^2}
\def\xdot{\dot{\!\ex}}
\def\y{\BM{y}}
\def\ys{\y^2}
\def\ydot{\dot{\!\y}}
\def\pat{\pa_t}
\def\pax{\pa_x}
\def\cia{C_{i\alpha}}
\def\cjb{C_{j\beta}}

\renewcommand{\thesection}{\arabic{section}}
\renewcommand{\theequation}{\thesection.\arabic{equation}}

\itp{96}{68}

\hfill{hep-th/9609190}\\

\vspace*{.2in}

\begin{center}
{\large\bf RENORMALIZING RECTANGLES AND OTHER TOPICS IN RANDOM MATRIX 
THEORY}
\end{center}

\begin{center}
{\bf Joshua Feinberg \& A. Zee}
\end{center}
\vskip 2mm
\begin{center}{{ Institute for Theoretical Physics}\\
{University of California,\\ Santa Barbara, CA 93106, USA}\\}
\end{center}
\vskip 3mm
\begin{abstract}
We consider random Hermitian matrices made of complex or real $M\times 
N$ rectangular blocks, where the blocks are drawn from various 
ensembles.
These matrices have $N$ pairs of opposite real nonvanishing eigenvalues, 
as 
well as $M-N$ zero eigenvalues (for $M>N$.) These zero eigenvalues are 
``kinematical" in the sense that they are independent of randomness. We 
study the eigenvalue distribution of these matrices to leading order in 
the large $N,M$ limit, in which the ``rectangularity" $r={M\over N}$ is 
held fixed. We apply a variety of methods 
in our study. We study Gaussian ensembles by a simple diagrammatic 
method, 
by the Dyson gas approach, and by a generalization of the Kazakov 
method. These methods make use of the invariance of such ensembles under 
the action of symmetry groups. The more complicated Wigner ensemble, 
which does not enjoy 
such symmetry properties, is studied by large $N$ renormalization 
techniques. 
In addition to the kinematical $\delta$-function spike in the eigenvalue 
density which corresponds to zero eigenvalues, we find for both 
types of ensembles that if $|r-1|$ is held fixed as 
$N\rightarrow\infty$, the $N$ non-zero eigenvalues give rise to two 
separated lobes that are located 
symmetrically with respect to the origin. This separation arises because 
the non-zero eigenvalues are repelled macroscopically from the origin. 
Finally, we study the oscillatory behavior of the eigenvalue 
distribution near the 
endpoints of the lobes, a behavior governed by Airy functions. As 
$r\rightarrow 1$ the lobes come closer, and the Airy oscillatory 
behavior near the endpoints that are close to zero breaks down. We 
interpret this breakdown as a signal that $r\rightarrow 1$ drives a 
cross over to the oscillation governed by Bessel functions near the 
origin for matrices made of square blocks.
\end{abstract}

\vspace{25pt}
PACS numbers: 11.10.Lm, 11.15.Pg, 11.10.Kk, 71.27.+a  
\vfill
\pagebreak

\setcounter{page}{1}

\section{Introduction}

In random matrix theory, a number of authors
\cite{Verbaarschot, amb, ambjorn, NS, Nforr, AST, bhznpb, Nishigaki} 
have
studied the eigenvalue distribution of a Hermitian matrix
$H$ of the form
\begin{equation}\label{C1}
  H =\left(\matrix{0& C^{\dag}\cr
                 C& 0 \cr}\right),
\end{equation}
in which $C$ is an  $N\times N$ complex random matrix taken from an ensemble with the
probability
distribution
\begin{equation}\label{Gaussian}
  P(C) = {1\over{Z}} {\exp} (-N {\rm Tr} C^{\dag}C)\,,
\end{equation}
with $N$ tending to
infinity. These so-called chiral matrices appear in a variety of physical 
problems.
For example, in quantum chromodynamics one typically integrates over the
quarks and studies the so-called fermion determinant. The gluon 
fluctuations
are then often treated approximately by saying that they effectively 
render
the relevant matrix in the determinant random \cite{Verbaarschot,
VerbaarschotZ, Zahed}. The chiral structure corresponds to left and 
right
handed quarks. As another example,
Hikami, Shirai, and Wegner \cite{HZ,HSW, hanna} have proposed a model 
for
electron scattering off impurities in quantum Hall fluids in the
spin-degenerate limit. The blocks in (1.1) correspond to spin up and 
spin
down electrons. In the same spirit, one may consider any problem 
involving
random scattering between two groups of states, for example, between two
cavities. As
pointed out by Nagao and Slevin \cite{NS}, these matrices
also appear in
the study of transport in disordered conductors.
In this paper, we study a slight generalization of this
problem, with $C$ taken to be an $M\times N$ rectangular matrix, with 
$M$
and $N$ both tending to infinity. For $M-N$ of order $N^0$, we expect 
the
density of eigenvalues to be the same as for the $M=N$ case. Here we 
would
like to study the case where the measure of rectangularity, 
\beq
r\equiv M/N\,,
\eeq
is held fixed as both $M$ and $N$ tend to infinity. Some aspects of this 
problem have been studied before and we will note the 
appropriate references below.
We denote the matrix elements of $C$ by 
\beqast
\cia, \quad {\rm where}\quad i=1, 2,~...,~M\quad {\rm and}\quad 
\alpha=1, 2,~...,~N. 
\eeqast
With no loss of generality we assume throughout
this paper that $M\geq N$, namely, that $r\geq 1$. Our notation is such 
that 
Latin indices always run from $1$ through $M$, whereas indices denoted 
by Greek letters run from $1$ to $N$.\footnote{We shall deviate slightly 
from this convention only in section $2$ where $\mu,\nu$ will run over 
all possible
$M+N$ values. No confusion should arise.} As a result of their specific  
structure these matrices have $N$ pairs of opposite real nonvanishing 
eigenvalues, as well as $M-N$ zero eigenvalues. These zero eigenvalues 
are ``kinematical" in the sense that they are independent of the 
probability distribution.

We derive the eigenvalue distribution of these matrices to leading order 
in the large $N,M$ approximation for various ensembles of random blocks. 
We consider random Hermitian matrices made of complex or real $M\times 
N$ rectangular blocks, where the blocks are drawn either from ensembles 
symmetric under some group action or from non-symmetric ensembles. For 
concreteness, we specialize to Gaussian ensembles in the first case. In 
the second case we analyze matrices of the ``Wigner Class", namely, 
blocks whose entries are drawn independently 
one of the other from the probability distribution. We find, not 
surprisingly, 
that to leading order in the large $N,M$ approximation, all the 
ensembles we studied result in the same eigenvalue distribution. In 
addition to the kinematical $\delta$-function spike in the eigenvalue 
density which corresponds to zero eigenvalues, we find that if $|r-1|$ 
does 
not scale to zero as $N\rightarrow\infty$, the $N$ non-zero eigenvalues 
give rise to two well separated lobes that are located symmetrically 
with respect to the origin. For random Hermitian matrices that are not 
made of blocks, the qualitative universality of the 
Wigner semicircular eigenvalue distribution is well understood as a 
result of the competition between level repulsion and the fact that very 
large eigenvalues are suppressed. Similar arguments explain the 
universality of the eigenvalue distribution we observe here
for matrices made of rectangular blocks. Each lobe arises qualitatively 
for the same reasons that lead to the semicircular distribution. In 
addition, separation of the two lobes arises because the non-zero 
eigenvalues are repelled from the origin by the macroscopic number $(M-
N)$ of zero eigenvalues.

In this paper $C^{\dgg}C$ and $CC^{\dgg}$ are Hermitian non-negative 
matrices of dimensions $N\times N$ and $M\times M$, respectively.
We are interested in the expectation value of the resolvent
\beq
\hat G_{N,M}(z) = {1\over N+M} ~\tr~ {1\over z-H}\,.
\label{gnm}
\eeq
A straightforward calculation then yields a simple relation between 
$\hat G_{N,M}(z)$ and the resolvents of $C^{\dgg}C$ and $CC^{\dgg}$, 
\beq\label{traceinverse}
\hat G_{N,M}(z) = {z\over N+M } 
\left[\trn {1\over z^2-C^{\dgg}C} + \trm {1\over z^2-CC^{\dgg}}\right]
\eeq
where the subscript on each trace indicates the dimension of the matrix 
which 
is being traced over.  The $z^2$ dependence of the resolvents in 
(\ref{traceinverse}) arises because the eigenvalues of $H$ in (\ref{C1}) 
occur in real opposite pairs. Indeed, given an $N$ dimensional vector 
$x$ and an $M$ dimensional 
vector $y$ such that 
 $\left(\matrix{x\cr y \cr}\right)$ is an eigenvector of $H$  for an 
eigenvalue
$\lambda $,  then  $\left(\matrix{x\cr -y \cr}\right)$ is an
eigenvector
for  $-\lambda $. In other words the matrix $H$ (the ``Dirac" operator, 
with 
its ``chiral" components $C$ and $C^{\dgg}$) anti-commutes with the 
$``\gamma_5"$ matrix $\left(\matrix{1_N& 0\cr
0& -1_M\cr}\right)$. The cyclic property of the trace implies the basic 
relation
\beq\label{mnrelation}
\trm {1\over z^2-CC^{\dgg}} = \trn {1\over z^2-C^{\dgg}C} + {M-N\over 
z^2}\,.
\eeq
This relation reflects the fact that $C^{\dgg}C$ and $CC^{\dgg}$ share 
the same
strictly positive eigenvalues, but the $M\times M$ matrix $\ccdg$ has 
additional $M-N$ zero eigenvalues.

Combining  (\ref{traceinverse}) and (\ref{mnrelation}) we therefore 
arrive
at the two alternative expressions
\beqra\label{simplegnm}
\hat G_{N,M}(z) &=& \left({M-N \over N+M}\right)~{1\over z} + {2z\over 
M+N}~\trn {1\over z^2-\cdgc}\nonumber\\
 &=& \left({N-M \over N+M}\right)~{1\over z} + {2z\over M+N}~\trm 
{1\over z^2-\ccdg}\,,
\eeqra
that allow us to express $\hat G_{N,M}(z)$ solely in terms of either 
$\cdgc$ 
or in terms of $\ccdg$. For later use we introduce the following 
notation
\beq\label{notation}
\hat G_N(w)={1\over N} \trn {1\over w-\cdgc}\,,\quad\quad \hat 
G_M(w)={1\over M} \trm {1\over w-\ccdg}
\eeq
in terms of which we rewrite (\ref{simplegnm}) as 
\beqra\label{simplergnm}
\hat G_{N,M}(z) &=& \left({M-N \over N+M}\right)~{1\over z} + 
\left({2N\over M+N}\right)~z\hat G_N(z^2)\nonumber\\
 &=& \left({N-M \over N+M}\right)~{1\over z} + \left({2M\over 
M+N}\right)~z\hat G_M(z^2)\,.
\eeqra
Throughout this paper $\hat G$ stands for an unaveraged resolvent. The 
corresponding averaged quantity will be denoted simply by $G$.

This paper is organized as follows. We will first apply a variety of
methods to study the density of eigenvalues. In Section 2 we derive the 
density of states of matrices $H$ whose rectangular blocks are drawn 
either from the unitary or from the orthogonal Gaussian ensemble, 
employing diagrammatic techniques. Section 3 is devoted to blocks with 
independent entries (which we refered to \cite{bzw} as the ``Wigner 
Class".)
This ensemble is more difficult to handle, because of lack of symmetry. 
We 
overcome this difficulty by applying recursive manipulations of the 
large $N$ renormalization group\cite{french, rg, daz}. We find that as 
far as
the density of states is concerned, this ensemble falls (in the
planar limit) into the same universality class as the symmetric 
ensembles.
In Appendix A we provide a proof of the central limit theorem by means 
of the large $N$ renormalization group, as yet another example of its 
usefulness. In Section 4 we present the Dyson gas approach to these 
issues. After 
completing our work we realized that the results we obtained following 
the 
Dyson gas approach already appeared in \cite{periwal}. Nevertheless, we 
include this section here for the paper to be self-contained and also 
because Section 5 partly relies on it. In Section 5 we first generalize 
Kazakov's method \cite{Kazakov} to rederive the results of Section 2, 
and then use this 
method to determine the oscillatory fine structure of the eigenvalue 
density in Section 2, close to its support endpoints. We find that 
this oscillatory 
behavior is governed by Airy functions. As $r\rightarrow 1$ the lobes 
come closer, and the Airy oscillatory behavior near the endpoints that 
are close to zero breaks down. We interpret this breakdown as a signal 
that in the limit $r\rightarrow 1$ drives a cross over to the oscillation near the 
origin in the density of eigenvalues of matrices made of square blocks, 
an oscillation governed by 
Bessel functions.

\pagebreak

\section{A diagrammatic approach}
\setcounter{equation}{0}

As a simple warm up exercise, and in order to set the stage, let us 
first
apply the by-now well-known diagrammatic method to derive the Green's 
function 
\beq\label{gzz}
G(z) = {1\over N+M} ~\langle\tr~ {1\over z-H}\rangle
\eeq
in the large $N,M$ limit. 
To this end, let us consider the averaged resolvent
\beq
G^{\mu}_{\nu}(z) = \langle\left({1\over z-H}\right)^{\mu}_{\nu}\rangle
\label{propagator}
\eeq
where the indices $\mu$ and $\nu$ run over all possible $M+N$ values.  
The average in (\ref{propagator}) is performed with respect to the 
Gaussian measure 
\beq\label{complexhermitean}
P(C)= {1\over{Z}} ~{\rm exp}~[-\sqrt{NM}~m^2~\tr~\cdgc]\,,
\eeq
where 
\beq
Z=\int \prod_{i=1}^M\prod_{\alpha = 1}^N d~{\rm Re}~C_{i\alpha} ~d~{\rm 
Im}~C_{i\alpha}~{\rm exp}~[-\sqrt{NM}~m^2~\tr~\cdgc]
\label{partition1}
\eeq
is the partition function. We have introduced a normalization factor of 
$\sqrt{MN}$ in (\ref{complexhermitean}) so as to be consistent with 
(\ref{Gaussian}) in the $N=M$ case. This factor renders 
(\ref{complexhermitean}) and (\ref{partition1}) manifestly symmetric 
under $M\leftrightarrow N$. Some other normalizations, not symmetrical 
under $M\leftrightarrow N$, can always be introduced by multiplying the 
parameter $m^2$ by an appropriate factor of $r={M\over N}$. Borrowing 
some terminology of gauge field theory we may consider $C, C^{\dgg}$ as 
``gluons" (in zero space-time dimensions), and $G^{\mu}_{\nu}(z)$ as the 
propagator of ``quarks" (with complex mass $z$) which couple to these 
``gluons". We now proceed to calculate $G^{\mu}_{\nu}(z)$ 
diagrammatically. 
The two-point correlator associated with (\ref{complexhermitean}) is 
clearly
\beq\label{correlatorhermitean}
\langle\cia\cjb^*\rangle\,=\,{1\over 
m^2\sqrt{MN}}\,\delta_{ij}\delta_{\alpha\beta}\,.
\eeq
This expression is the gluon propagator. The bare quark propagator is 
simply ${1\over z}$. The quark-quark-gluon vertex factor is $1$. These 
Feynman rules are summarized in Fig. (1). 

\vspace{36pt}
\begin{center}
\epsfbox{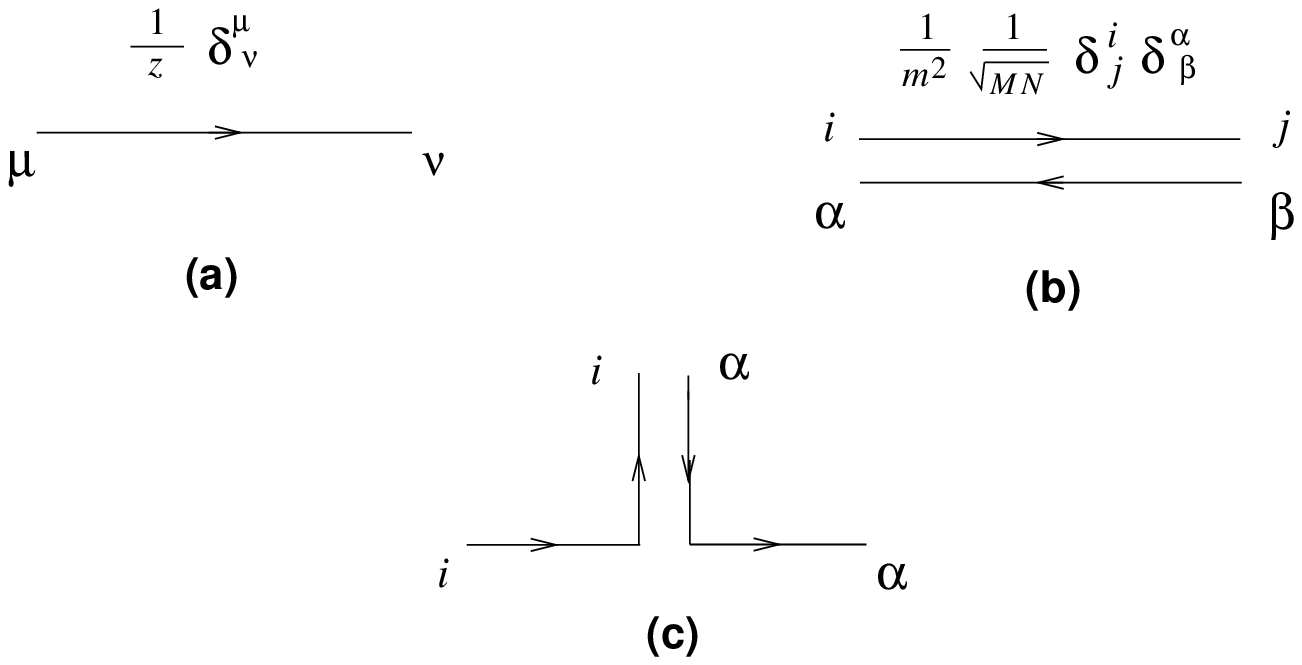}
\end{center}
{\footnotesize {\bf Fig. (1) - } The Feynman rules: (a) The bare quark propagator, (b) 
The gluon propagtor, (c) The bare quark-quark-gluon vertex.}
\vspace{36pt} 

The weight (\ref{complexhermitean}) is Gaussian, so there are no gluonic 
self interactions. Due to the block structure of $H$, the quark-quark-
gluon vertex converts the color type carried by a quark. Namely, it 
converts a quark $q_i$ into a quark $q_{\alpha}$ (and vice-versa), but 
not into a quark $q_j$. Recall that the dominating diagrams in the large 
$N,M$ limit are all planar, and
thus do not contain crossed color lines. Moreover, the (single) fermion 
line 
must always be at the boundary spanned by the planar graph. Consider now 
a typical planar
Feynman diagram in the perturbative expansion of (\ref{propagator}). 
Tracking the color indices through the diagram, we see that the 
rectangular off-diagonal blocks of (\ref{propagator}) vanish identically
\beq
G^{\alpha}_{i}(z) = G^{i}_{\alpha}(z) = 0, 
\label{offdiagonalblocks}
\eeq
while the diagonal blocks are proportional to unit matrices,
\beq
G^{i}_{j}(z) =  g_M (z)~ \del^i_j ,\quad\quad G^{\alpha}_{\beta}(z) = 
g_N (z) 
~\del^{\alpha}_{\beta}\,.
\label{diagonalblocks}
\eeq
Upon comparison with (\ref{notation}) we clearly have 
\beq\label{gG}
g_N(z)=zG_N(z^2)\quad\quad{\rm  and}\quad\quad g_M(z)=zG_M(z^2)\,.
\eeq
Figure (2) shows some of the diagrams which contribute to $G^i_j (z)$ to 
leading order in large $M,N$.

\vspace{36pt}
\begin{center}
\epsfbox{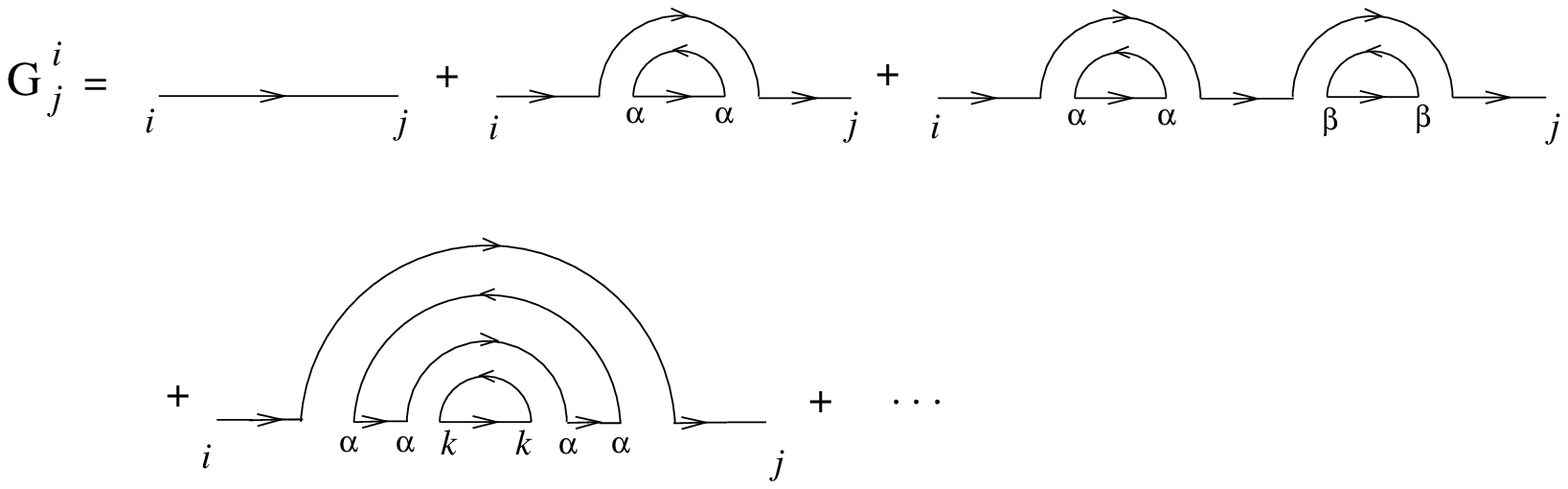}
\end{center}
{\footnotesize {\bf Fig. (2) - } The first few planar diagrams that contribute to 
$G^{i}_{j}$ to leading order in the large $N,M$ approximation.} 
\vspace{36pt}

A simple direct calculation shows that the off diagonal blocks of $1/(z-
H)$ are odd in $C, C^{\dgg}$ and thus do not contribute to 
(\ref{propagator}), independently of the perturbation expansion. This 
conclusion
clearly remains valid if we generalized (\ref{complexhermitean}) into 
any other
probability distribution which is even in $H$, for example, a 
distribution of the form $P(C)={1\over Z} {\rm exp}~ \left[-\tr V(\cdgc) 
\right]$. The self-energy corrections $\Sigma_N(z)$ 
and $\Sigma_M(z)$ are 
defined as usual by
\beq\label{selfenergy}
g_N(z)={1\over z- \Sigma_N(z)}\,,\quad\quad g_M(z)={1\over z- 
\Sigma_M(z)}
\eeq
and correspond to the sum over all one quark irreducible graphs 
contributing
to (\ref{propagator}), namely, the amputated quark propagator. For the 
Gaussian distribution (\ref{complexhermitean}), the propagators
$g_N, g_M$ and self-energies $\Sigma_N, \Sigma_M$ are related by the 
simple Schwinger-Dyson identities which we display diagrammatically on 
Fig. (3).

\vspace{36pt}
\begin{center}
\epsfbox{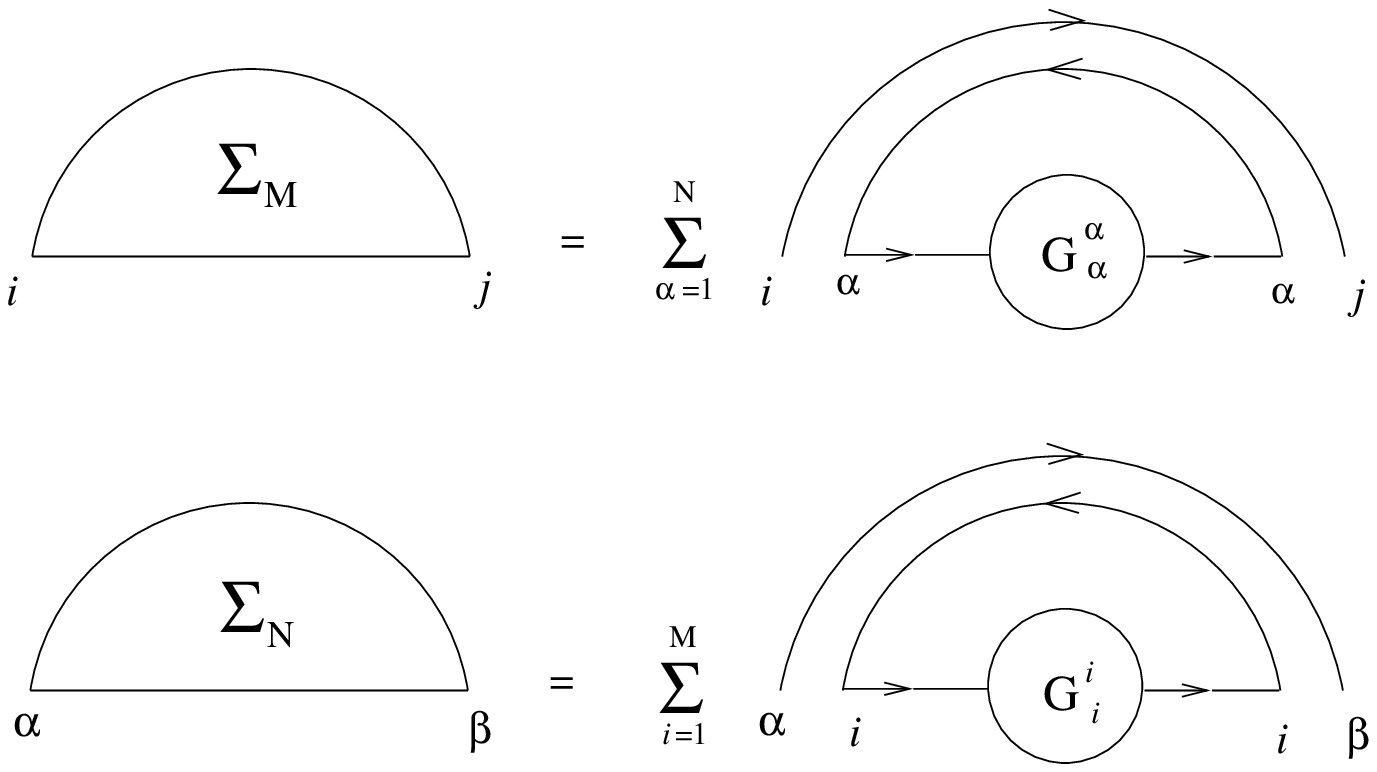}
\end{center}
{\footnotesize {\bf Fig. (3) - } The two Schwinger-Dyson identities.}
\vspace{36pt}

We thus have from Figs. (1) and (3)
\beq\label{sigmangm}
\Sigma_N(z) = {1\over m^2\sqrt{NM}}\sum_{i = 1}^M G^i_i(z) = 
{\sqrt{r}\over m^2}
g_M(z)
\eeq
and similarly,
\beq\label{sigmamgn}
\Sigma_M(z) = {1\over m^2 \sqrt{r}} g_N(z)\,.
\eeq
We substitute the last two equations into (\ref{selfenergy}) and obtain
the two coupled equations
\beq\label{coupled}
\left\{\begin{array}{c} g_N(z)=\left[ z- {\sqrt{r}\over 
m^2}g_M(z)\right]^{-1}\\{}\\
g_M(z)=\left[ z- {1\over m^2 \sqrt{r}} g_N(z)\right]^{-
1}\end{array}\right.
\eeq
for $g_N$ and $g_M$. These two equations clearly transform one into the 
other under $r\rightarrow {1\over r}$, which interchanges $g_N$ and 
$g_M$ (this is as it should be if the normalization factor in 
(\ref{complexhermitean}) is symmetrical under $M\leftrightarrow N$, that 
is, if $m^2$ is independent of $r$.)  By definition, both 
propagators behave as ${1\over z}$ in the asymptotic region 
$z\rightarrow\infty$. This asymptotic behavior picks up the physical 
solution
of the quadratic equations for $g_N$ and $g_M$, and we find
\beqra\label{gngm}
g_N(z) &=& {2\over\left( a-b\right)^2}~{1\over z}~\left[z^2-ab - 
\sqrt{\left(z^2-a^2\right)\left(z^2-
b^2\right)}\right]\nonumber\\{}\nonumber\\
g_M(z) &=& {2\over\left(a+b\right)^2}~{1\over z}~\left[z^2+ab - 
\sqrt{\left(z^2-a^2\right)\left(z^2-b^2\right)}\right],
\eeqra
where 
\beq
a = {1\over m}~(r^{1\over 4}+r^{-{1\over 4}})\quad\quad,\quad\quad b = 
{1\over m}~(r^{1\over 4}-r^{-{1\over 4}}). 
\label{ab}
\eeq
Note that $b$ measures of the deviation of rectangles from squares: for 
$r=1$, $b$ vanishes. The Green's function (\ref{gzz}) is thus given by
\beq
G(z) = {1\over N+M} \sum_{\mu=1}^{N+M} G_{\mu}^{\mu} (z) = {g_N(z)+ r 
g_M(z)\over 1+r}
\label{gdiagram}
\eeq
where we used (\ref{propagator}) and (\ref{diagonalblocks}). Finally, 
upon substituting (\ref{gngm}) into (\ref{gdiagram}) we find that the  
averaged Green's function of $H$ is
\beq
G(z) = {2\over a^2+b^2}~{1\over z}~\left[ z^2 - \sqrt{(z^2-b^2)(z^2-
a^2)}\right]\,.
\label{gzhermitean}
\eeq

Let us inspect now some of the features of (\ref{gzhermitean}). As we 
discussed in the introduction, $H$ has $M-N$ ``kinematical" zero 
eigenvalues, regardless of any ensemble averaging. In contrast, $\cdgc$ 
on the average does not have any zero eigenvalues as we have already 
discussed. Thus, by definition, $G_{N,M}(z)$ has a simple pole at $z=0$ 
with residue ${M-N\over M+N} = {r-1\over r+1}$, which is the first term 
on the right side of (\ref{simplergnm}). As we can see from (\ref{ab}), 
our expression (\ref{gzhermitean}) clearly satisfies this condition 
provided $\sqrt{(z^2-b^2)(z^2-a^2)}\rightarrow -ab$ as $z\rightarrow 0$. 
This sign of the square root corresponds to drawing all four cuts 
associated with the square root to the left of the branch point out of 
which they emanate. In addition, (\ref{ab}) is consistent with the 
required
asymptotic behavior ${1\over z}$ of (\ref{gzhermitean}) as 
$z\rightarrow\infty$. The averaged eigenvalue density
of $H$ is the discontinuity in (\ref{gzhermitean}) as we cross the real 
axis, except for the origin, which contains the ``kinematical" zero 
eigenvalues of $H$. It is therefore given by 
\beq\label{rhohermitean}
\rho(\lambda) =  {r-1\over r+1}~ \del(\lambda) + {2\over \pi |\lambda|}~ 
{\theta\left[\left(a^2-\lambda^2\right)\left(\lambda^2-
b^2\right)\right]\over a^2+b^2}~\sqrt{\left(a^2-
\lambda^2\right)\left(\lambda^2-b^2\right)}\,.
\eeq

The Green's function (\ref{gzhermitean}) corresponds to the Hamiltonian 
(\ref{C1}). With only little more effort it is possible to generalize 
our discussion to calculate the Green's function of the Hamiltonian
\begin{equation}\label{He}
  H =\left(\matrix{\epsilon & C^{\dag}\cr
                 C& -\epsilon \cr}\right),
\end{equation}
where $\epsilon$ is a fixed ``energy". The off-diagonal blocks fluctuate 
as before. This Hamiltonian may describe, for example, tunneling between two 
energy levels with degeneracies $N$ and $M$ that are seperated by an 
energy difference $2\epsilon$. For such a Hamiltonian 
(\ref{traceinverse}) is modified into 
\beq\label{nondegenerategnm}
G_{N,M}(z) = {z+\epsilon\over r+1} G_N(w) +  {r \left(z-
\epsilon\right)\over r+1} G_M(w)
\eeq
where 
\beqast
w=z^2-\epsilon^2\,,
\eeqast 
such that the identifications (\ref{gG}) become
\beq\label{gG1}
g_N(z)=(z+\epsilon) G_N(w)\quad\quad{\rm  and}\quad\quad g_M(z)=(z-
\epsilon)G_M(w)\,.
\eeq
The bare quark propagator in Fig. (1) is split into two pieces, namely, 
${1\over z-\epsilon}$ for quarks carrying a $U(N)$ color index and 
${1\over z+\epsilon}$ for quarks carrying a $U(M)$ index. 
The definitions in (\ref{selfenergy}) change accordingly into 
\beq\label{selfenergynondeg}
g_N(z)={1\over z-\epsilon -  \Sigma_N(z)}\,,\quad\quad g_M(z)={1\over z 
+ \epsilon - \Sigma_M(z)}\,.
\eeq
The Schwinger-Dyson identities (\ref{sigmangm})
and (\ref{sigmamgn}) are unchanged. We note that the set of equations 
(\ref{sigmangm}), (\ref{sigmamgn}) and (\ref{selfenergynondeg}) are now 
invariant under $r\rightarrow {1\over r}$ and $\epsilon\rightarrow 
-\epsilon$, which permutes the two energy levels in $H$ and thus 
interchanges $g_N$ and $g_M$. With these observation it is straightforward to see that (\ref{gngm}) becomes 
\beqra\label{gngm1}
g_N(z) &=& {2\over \left( a-b\right)^2}~{1\over \left(z-
\epsilon\right)}~\left[w-ab - \sqrt{\left(w-a^2\right)\left(w-
b^2\right)}\right]\nonumber\\{}\nonumber\\
g_M(z) &=& 
{2\over\left(a+b\right)^2}~{1\over\left(z+\epsilon\right)}~\left[w+ab - 
\sqrt{\left(w-a^2\right)\left(w-b^2\right)}\right],
\eeqra
with the same $a$ and $b$ as before. Thus, (\ref{gdiagram}) and 
(\ref{gzhermitean}) finally become
\beq
G(z) = {2\over \left(a^2+b^2\right)}~{1\over w}~ \left\{z\left[w-
\sqrt{\left(w-a^2\right)\left(w-b^2\right)}\right] - a b \epsilon 
\right\}\,,
\label{gdiagram1}
\eeq 
which is manifestly invariant under a permutation of the two energy 
levels of
$H$. Note that the matrix (\ref{He}) has precisely $M-N$ ``kinematical" 
({\em i.e.}, independently of the ensemble for $C$) eigenvectors which 
correspond to eigenvalue $-\epsilon$. This means that (\ref{gdiagram1}) 
has a simple pole at 
$z=-\epsilon$ with residue ${r-1\over r+1}$, and no pole at 
$z=+\epsilon$.
This property holds provided $\sqrt{\left(w-a^2\right)\left(w-
b^2\right)}\rightarrow -ab$ as $w\rightarrow 0$ which we already 
encountered in our analysis of the $\epsilon=0$ case.
The eigenvalue distribution corresponding to (\ref{gdiagram1}) is 
therefore
\beqra\label{rhoepsilon}
\!\!\!\!\!&&\rho(\lambda) =  {r-1\over r+1}~ \del(\lambda + \epsilon) + 
\nonumber\\{}\nonumber\\&&{\theta\left[\left(a^2 + \epsilon^2 
-\lambda^2\right)\left(\lambda^2-b^2 - \epsilon^2\right)\right]\over 
\pi\left(a^2+b^2\right)}{2~|\lambda |\over\left(\lambda^2 - 
\epsilon^2\right)}\sqrt{\left(a^2+\epsilon^2-
\lambda^2\right)\left(\lambda^2-b^2-
\epsilon^2\right)}\,.
\eeqra

In the limit $(m\epsilon)\rightarrow\infty$, the randomness in 
(\ref{He}) is suppressed, and (\ref{rhoepsilon}) should reproduce the 
eigenvalue distribution of the deterministic part of (\ref{He}). This is 
indeed the case. In this limit we have 
${a\over\epsilon},\,{b\over\epsilon}\rightarrow 0$ so both lobes in 
(\ref{rhoepsilon}) shrink. Each lobe contains $N$ eigenvalues, whose 
number is preserved as the lobes shrink to zero width, and thus produce 
$\delta$ function spikes of strength ${N\over M+N}$ each. The right lobe 
produces in this way a spike at $\lambda=\epsilon$, while the left lobe 
coalesces with the already existing $\delta (\lambda + \epsilon)$ spike 
in (\ref{rhoepsilon}) which contains $M-N$ eigenvalues, and thus produces a 
spike containing $M$ eigenvalues.

\pagebreak

\section{Blocks with independent matrix elements and their 
renormalization group analysis}
\setcounter{equation}{0}

It is rather difficult to apply the diagrammatic method and sum all the 
planar diagrams that contribute to $G(z)$ in case of non-Gaussian 
ensembles. For such
ensembles that are invariant under the action of some symmetry group one 
may
invoke other methods. However, these methods are inapplicable to 
ensembles lacking the action of a symmetry group. 

A class of block structure random matrix models that is not unitary (or 
orthogonal) invariant involves matrix blocks $C$ in which each matrix 
element $\cia$ is randomly distributed independently of the others, with 
the same distribution for all matrix elements. We normalize the matrix 
elements $\cia$ symmetrically with respect to $M$ and $N$, such that the 
two-point correlator 
\beq\label{correlator}
\langle\cia\cjb^*\rangle\,=\,{\si^2\over\sqrt{MN}}\delta_{ij}
\delta_{\alpha\beta}
\eeq
of this probability distribution would coincide with the two point 
correlator 
of the Gaussian distribution $d\mu(C)\sim {\rm exp}\left[-
{\sqrt{MN}\over \si^2}\tr\cdgc\right]$. For notational simplicity we 
replaced here the $m^{-2}$ in (\ref{complexhermitean}) by $\si^2$. For 
this class of matrix models the 
method of orthogonal polynomials is not directly applicable, and 
alternative methods should be sought for.

For concreteness as well as for simplicity, we consider below the 
probability distribution in which $\cia$ may take one of the two values
\beq\label{cia}
\pm{\si\over\left(NM\right)^{1/4}}
\eeq
with equal probability, where $\si$ is a finite number. However, it will 
be clear from the discussion below, that our conclusions are not limited 
to this particular distribution. In order to keep our formulas generic, 
we therefore treat the $\cia$ as complex numbers, as long as we do not 
utilize (\ref{cia}) explicitly.

For this ensemble $|C_{i\alpha}|^2={\si^2\over\sqrt{MN}}$ 
deterministically, and thus in particular the diagonal matrix elements 
of $\cdgc$ and $\ccdg$ do not fluctuate and are given by
\beq
(\cdgc)_{\alpha\alpha}=\si^2\sqrt{r}\quad {\rm and}\quad 
(\ccdg)_{ii}=\si^2/\sqrt{r}\,.
\label{diagonalelements}
\eeq
We use this convenient property of (\ref{cia}) in our calculations 
below. This, however, is done with no loss of generality, because in the 
generic case the 
non-fluctuating quantities in (\ref{diagonalelements}) should simply be 
replaced by their averages, which are given on the right hand side of  
(\ref{diagonalelements}).

The random matrix distribution we 
consider is a generalization of the very first model studied by 
Wigner\cite{wigner} into random matrices with block structure. Indeed, 
Wigner originally considered large $N\times N$ random Hermitean matrices 
$\phi$, whose elements $\phi_{ij}=\phi_{ji}^{*}$ were either $+{\si\over 
\sqrt{N}}$ ~or $-{\si\over \sqrt{N}}$, with equal probability. This 
matrix model follows a semi-circle law for the density of eigenvalues.  
This semi-circular profile of the eigenvalue density was rederived 
recently\cite{french} using a large N ``renormalization group" inspired 
approach\cite{rg}. In what follows, we apply
the same method to find the eigenvalue density $\rho(\lambda)$ of the 
random block matrices with independent entries introduced above.

We are interested in 
\beq\label{gz}
G(z)=\nmasymptotic G_{N,M}(z)
\eeq
from which $\rho(\lambda)={1\over\pi}{\rm Im}G(\lambda - i\epsilon)$ may 
be 
extracted immediately. We start our ``renormalization group" calculation 
by
trying to relate $G_{N+1,M}(z)$ to $G_{N,M}(z)$. To this end we consider 
the $M\times (N+1)$ block
\beqast
C'=(C, v)
\eeqast
where $v$ is an $M$ dimensional vector. By definition, each element of 
$C'$ may take now one of the two values 
$\pm{\si\over\left[\left(N+1\right)M\right]^{1/4}}$ with equal 
probability.
A comparison with the original $N\times M$ block suggests then that we 
may draw the $\cia'$ from (\ref{cia}) provided we rescale the $\si$ 
parameter in that equation into 
\beq\label{sigmaprime}
\si'=\si\left({N\over N+1}\right)^{1\over 4}\,.
\eeq
The non-fluctuating norm squared of $v$ is then given by 
\beq\label{vsquare}
\vdgv\equiv(\pcdgc)_{N+1,N+1}=\si'^{~2}\sqrt{r}\,.
\eeq
Following \cite{french}, we obtain\footnote{The analogue of $\vdgv$ in 
\cite{french} was a quantity of $\cO({1\over N})$ which was therefore 
neglected
in the $N\rightarrow\infty$ limit. Here $\vdgv$ is a finite number and 
must be therefore retained.} after some straightforward algebra 
\beqra\label{recursion}
(N&+&1)\hat G_{N+1}(w) \equiv \trnn {1\over w-\pcdgc}\nonumber\\
 &=& \trn {1\over w-\cdgc -C^{\dgg}{v\otimes v^{\dgg} \over w-\vdgv}C} +  
{1\over w-\vdgv -v^{\dgg}C {1 \over w-\cdgc}C^{\dgg}
v}  
\eeqra
where $w=z^2$.
We now average over the distribution governing $C'$, keeping terms up
to $\cO(N^0)$. We first average over the components of $v$.

Expanding the two fractions on the right hand side of (\ref{recursion}) 
into
geometric series we see that we have to average products of the form 
$v_i^*v_jv_k^*v_l\cdots v_p^*v_q$, which contract against products of 
elements of matrices independent of the $v_i$ . 
Clearly,
\beqast
<v_i^* 
v_j>=<C_{i,N+1}^{~'*}C'_{j,N+1}>={\si'^{~2}\over\sqrt{MN}}\delta_{ij}
\eeqast
simply produces a single trace, multiplied by ${\si'^{~2}\over 
\sqrt{NM}}$. The next non-vanishing correlator is
\beq
<v_i^*v_jv_k^*v_l>={\si'^{~4}\over 
NM}(\delta_{ij}\delta_{kl}+\delta_{ik}\delta_{jl}-\delta_{ijkl})\,,
\label{fourth}
\eeq
where $\delta_{ijkl}=1$ when $i=j=k=l$ and $0$ otherwise. The last piece 
in (\ref{fourth}) is by definition the fourth order cumulant of the 
distribution, added to the usual pairs of Wick contractions. The 
correlator (\ref{fourth}) then contracts against two matrices in the 
geometrical series mentioned above, producing a term of order 
$\left({\si'^{~2}\over \sqrt{NM}}\right)^2$. In the large $N,M$ limit, 
the dominating term in this contraction is the term with the maximal 
independent index summations, which amounts here to two traces. These 
two traces are produced here only by a single pair of
Wick contractions. The other pair of Wick contractions (which produces 
only a single trace) as well as the fourth order cumulant are therefore 
negligible in the large $N,M$ limit, and may be discarded to leading 
order.
This structure persists for correlators of higher order. The $2n$ point 
correlator produces in the geometric series a term proportional to 
$\left({\si'^{~2}\over \sqrt{NM}}\right)^n$. In that term, a unique 
string
of $n$ Wick contractions  produces the maximal number $n$ of traces, and 
therefore dominates the large $N,M$ limit. At this point it becomes 
clear 
why our calculation, and therefore, the results it leads to are 
insensitive
to the details of the distribution of the $\cia$. Clearly, only Wick 
contractions dominate these averages in this limit, and thus only the 
two 
point function (\ref{correlator}) of the distribution matters. This 
insensitivity to the detials of the distribution was checked explicitly 
in \cite{french}, where various distributions led to the same result.
The leading terms in the geometric series may be resummed, and one finds 
the
$v$ average
\beqra\label{recursionv}
&&\langle\trnn {1\over w-\pcdgc}\rangle_v = \trn {1\over w-
\cdgc}\hfill\nonumber\\{}\nonumber\\
&+& {\pa\over \pa w} {\rm log} \left[w+(N-M){\si'^{~2}\over \sqrt{NM}}-
{\si'^{~2}\over \sqrt{NM}} ~w~\trn {1\over w-\cdgc}\right]\,.
\eeqra
Invoking large $N$ factorisation, we can average over the remaining 
block $C$ 
immediately, by replacing $\hat G_N$ inside the logarithm in 
(\ref{recursionv}) by
its average. Thus,
\beq\label{recursiongn}
(N+1)G_{N+1}(w)=NG_N(w)+{\pa\over \pa w} {\rm log} \left[w+{(1-
r)\si'^{~2}\over \sqrt{r}}-{\si'^{~2}\over \sqrt{r}} ~w~G_N(w)\right]\,.
\eeq
Remarkably, in the large $N,M$ limit, the $v$ average of $\hat 
G_{N+1}(w)$ involves only $\hat G_{N}(w)$, and thus (\ref{recursiongn}) 
is indeed a local (along the $N$ axis) recursion relation involving only 
$G_N$ type Green's functions. This means that the large $N$ 
``renormalization group" recursions
for the full Green's function $G_{N,M}$ close among themselves as we now 
show.

Combining (\ref{simplergnm}) and (\ref{recursiongn}), we obtain the
recursion relation for the complete averaged Green's function 
(\ref{gnm})
\beqra\label{recursiongnm}
&&(N+M+1)G_{N+1,M}(z,\si')-(N+M)G_{N,M}(z,\si')=\nonumber\\{}\nonumber\\
&&{\pa\over \pa z} {\rm log} \left[z+{(1-r)\si'^{~2}\over 2z\sqrt{r}}-
{(r+1)\si'^{~2}\over 2\sqrt{r}} G_{N,M}(z,\si')\right]\,,
\eeqra
where we have displayed the explicit $\si'$ parameter associated with 
the 
larger $C'$ block.

In the large $N$ limit (\ref{sigmaprime}) becomes $\si'=\si-{\si\over 
4N} +\cdots$. In this limit, the only possible explicit $N,M$ dependence 
in $G_{N,M}$ is through the finite ratio $r={M\over N}$. Therefore we 
may write 
the left hand side of (\ref{recursiongnm}) as
\beqra\label{lhs}
&&\left[(N+M+1)G_{N+1,M}(z,\si')-(N+M)G_{N,M}(z,\si)\right]+(N+M)\left[         
G_{N,M}(z,\si)-G_{N,M}(z,\si')\right]\nonumber\\{}\nonumber\\
&=&{\pa\over \pa N}\left[(N+M)G_{N,M}(z,\si)\right]+{(N+M)\over 
4N} \si{\pa\over \pa\si}G_{N,M}(z,\si)\nonumber\\{}\nonumber\\
&=&G_{N,M}(z,\si)-r(r+1){\pa\over\pa r}G_{N,M}(z,\si)+{r+1\over 4}\si 
{\pa\over\pa\si}G_{N,M}(z,\si)\,.\eeqra
To leading order in ${1\over N}$ we may drop the $N,M$ indices of the 
Green's function, replacing it by its asymptotic limit (\ref{gz}), and 
replace $\si'$ by $\si$ inside the logarithm in (\ref{recursiongnm}). 
The recursion relation
(\ref{recursiongnm}) thus becomes a partial differential equation
\beqra\label{diffeq}
&&G(z,\si)-r(r+1){\pa\over\pa r}G(z,\si)+{r+1\over 4}\si 
{\pa\over\pa\si}G(z,\si)=\nonumber\\{}\nonumber\\
&&{\pa\over \pa z} {\rm log} \left[z+{(1-r)\si^2\over 2z\sqrt{r}}-
{(r+1)\si^2\over 2\sqrt{r}} G(z,\si)\right]\,
\eeqra

It is easy to see from (\ref{gnm}) and (\ref{gz}) that $G(z,\si)$ 
satisfies the
simple scaling rule
\beq\label{scaling}
G\left(z,\si\right)={1\over\si}G\left({z\over\si},1\right)
\eeq
which implies that 
\beq\label{scaling1}
\si{\pa\over \pa\si}G(z,\si)=-z{\pa\over \pa z}G(z,\si) - G(z,\si)\,.
\eeq
Thus, using (\ref{scaling1}) to eliminate $\si{\pa\over \pa\si}G(z,\si)$
from (\ref{diffeq}) we arrive at the final form of our differential 
equation
for $G(z,\si)$, namely,
\beqra\label{diffeqfinal}
&&{3-r\over 4}G(z,\si)-r(r+1){\pa\over\pa r}G(z,\si)-{r+1\over 4}z 
{\pa\over\pa z}G(z,\si)=\nonumber\\{}\nonumber\\
&&{\pa\over \pa z} {\rm log} \left[z+{(1-r)\si^2\over 2z\sqrt{r}}-
{(r+1)\si^2\over 2\sqrt{r}} G(z,\si)\right]\,.
\eeqra
This equation tells us how a change in $z$ can be compensated by a 
change in 
the rectangularity $r$.

As a consistency check of our results we can repeat the recursive 
procedure
discussed above, but instead of adding an $M$ dimensional column vector 
to $C$, we add to it an $N$ dimensional row vector $u$, creating an 
$(M+1)\times N$ block $C''$
\beqast
C''=\left(\begin{array}{c} C\\{}\\ u\end{array}\right)\,.
\eeqast
The recursion relation in this case connects, in the large $N,M$ limit,
$G_{M+1}(w)$ and $G_M(w)$, and therefore relates $G_{N,M+1}(z)$ to 
$G_{N,M}(z)$. Thus, we simply interchange $N\leftrightarrow M$ in all 
steps of 
our calculation above, namely, $r\leftrightarrow {1\over r}$. The 
differential
equation for $G(z,\si)$ we derived from this recursion reads
\beqra
&&{3r-1\over 4r}G(z,\si)+(r+1){\pa\over\pa r}G(z,\si)-{r+1\over 4r}z 
{\pa\over\pa z}G(z,\si)=\nonumber\\{}\nonumber\\
&&{\pa\over \pa z} {\rm log} \left[z-{(1-r)\si^2\over 2z\sqrt{r}}-
{(r+1)\si^2\over 2\sqrt{r}} G(z,\si)\right]\,,
\eeqra
which is indeed the transform of (\ref{diffeqfinal}) under $r\rightarrow 
{1\over r}$.

The fact that $G(z,\si)$ satisfies both (\ref{diffeqfinal}) and its 
tranform under $r\rightarrow {1\over r}$ means that $G(z,\si,r)=G(z,\si, 
{1\over r})$. This $r$ inversion 
symmetry of $G$ should be anticipated from our $N,M$ symmetric 
definition
of the probability distribution (\ref{correlator}) in the first place. 
An important consequence of this $r$ inversion symmetry is that 
${\pa\over \pa r} G$ vanishes\footnote{This is simply because if 
$f(r)=f\left(r^{-1}\right)$, then ${\pa\over \pa r} f(r) = - r^{-
2}{\pa\over \pa r^{-1}}f\left(r^{-1}\right)$, and therefore $f'(1)=-f'
(1)=0$.}at $r=1$. Thus, at the point $r=1$, {\em i.e.}, for 
Hamiltonians made 
of square blocks, (\ref{diffeqfinal}) reduces to the differential 
equation 
\beq
G(z,\si)-z {\pa\over\pa z}G(z,\si) = 2{\pa\over \pa z} {\rm log} 
\left[z-\si^2~ G(z,\si)\right]
\eeq
previously derived in \cite{french}, as expected.

As was stated at the beginning of this section, only the two point 
correlator (\ref{correlator}) of the random matrix distribution was 
relevant in the derivation of (\ref{diffeqfinal}). Hence, the Green's 
function $G(z)$ of any distribution obeying (\ref{correlator}) is a 
solution of (\ref{diffeqfinal}). 
We have thus shown that for the Wigner ensemble $G(z)$ and the density 
of 
eigenvalues are universal. In particular, the complex Hermitean 
distribution (\ref{complexhermitean}) as well as the real symmetric 
distribution (\ref{realsymmetric}) of the previous section respect 
(\ref{correlator}) upon the identification $m^2=\si^{-2}$. Thus, their 
Green's function (\ref{gzhermitean}) must be a solution of 
(\ref{diffeqfinal}). A simple check verifies that this is indeed the 
case.
Therefore, the density of eigenvalues $\rho(\lambda)={1\over\pi}{\rm 
Im}G(\lambda-i\epsilon)$ is given by
(\ref{rhohermitean}).

As yet another example of the usefulness of the large $N$ 
renormalization group we use it to prove the central limit 
theorem in Appendix A.

By a simple power counting argument (see Section 2 of \cite{bzw}, and 
also \cite{daz}) it is straightforward to extend the diagrammatic 
method of the previous 
section to treat the probability distribution considered in this section 
as well.

\pagebreak

\section{Dyson gas approach}
\setcounter{equation}{0}

In this section we present the Dyson gas approach to study the 
eigenvalue distribution of matrices made of rectangular blocks. After 
completing our calculations we realized that our results were previously 
obtained by Periwal et al. in \cite{periwal}.  We assume that the 
$M\times N$ rectangular blocks 
$C_{i\alpha}$ of the Hamiltonian $H$ in (\ref{C1}) admit the action of 
some symmetry group. Here we focus on blocks with complex entries,
but we will state some results concerning blocks with real entries in 
the end. The complex blocks are endowed with the natural 
$U(M)\times U(N)$ action
\beq
C\rightarrow V C U\,,\quad\quad\quad V\in U(M)\,,\quad U\in U(N)\,.
\label{umn}
\eeq
One can use this action to bring $C$ to the form 
\beq
C= \left(\begin{array}{c} \Lambda_N\\{}\\ 0_{\left(M-N\right)\times 
N}\end{array}\right)
\label{form}
\eeq
where $\Lambda_N$ is a real diagonal $N\times N$ matrix ${\rm 
diag}(\lambda_1, \cdots, \lambda_N)$. Therefore, the Hermitian matrix 
$H$ in (\ref{C1}) is a generator of the symmetric space $U(M+N)/ 
U(M)\otimes U(N)$. From these considerations it is clear that $\cdgc$ 
may be diagonalized into ~${\rm diag}(\lambda_1^2, \cdots,\lambda_N^2)$~ 
and $\ccdg$ into the same form, but
with additional $M-N$ zeros, in accordance with (\ref{mnrelation}). The 
probability distribution has to be invariant under (\ref{umn}). Here we 
consider distributions of the form 
\beq\label{pc}
P(C) = {1\over Z} ~ {\rm exp}~ \left[ -\sqrt{MN}~\tr ~ V(\cdgc)\right]
\eeq
where $V$ is a polynomial and $Z$ is the partition function of 
these matrices. 

We are interested only in averages of quantities that are invariant 
under (\ref{umn}). We thus transform from the Cartesian coordinates 
$C_{i\alpha}$ to polar coordinates $V_{ij}, U_{\alpha\beta}$ and 
$\lambda_{\alpha}$. Integrations over the unitary groups are irrelevant 
in calculating averages of invariant quantities, which involve only the 
eigenvalues $s_{\alpha}=\lambda_{\alpha}^2$ of $\cdgc$.

The partition function for these eigenvalues then reads \cite{periwal}
\beq
Z =\prod_{\alpha=1}^{N}~ \int\limits_{0}^{\infty}d s_{\alpha} ~{\rm 
exp}~[-\sqrt{NM}~V(s_{\alpha}) ]  \prod_{\beta=1}^{N} s_{\beta}^{M-N} 
\prod_{1\leq\gam<\del\leq N} (s_{\gam}-s_{\del})^2 \,.
\label{partition}
\eeq
The last two products constitute the Jacobian associated with polar 
coordinates. In particular, $\prod (s_{\gam}-s_{\del})^2$ is the 
familiar Vandermonde determinant. The other product is a feature 
peculiar to non-square blocks. As a trivial check of the validity of 
(\ref{partition}), note that the integration measures in 
(\ref{partition1}) and (\ref{partition}) have the same scaling dimension 
under $C\rightarrow \xi C, \quad \xi>0$.

Following Dyson, we observe that (\ref{partition}) may be interpreted as 
the partition function for a one dimensional gas of particles whose 
coordinates 
are given by the eigenvalues $s_{\alpha}$. The integrand in 
(\ref{partition}) may be expressed as ${\rm exp}~[-\sqrt{NM}~{\cal E}] $ 
where 
\beq
{\cal E} = \sum_{\alpha = 1}^N \left[ V (s_{\alpha})  - {r-1 \over 
\sqrt{r}}~{\rm log} ~s_{\alpha}\right] - {1\over 
N\sqrt{r}}\sum_{1\leq\alpha<\beta\leq N} {\rm log}~ (s_{\alpha}-
s_{\beta})^2
\label{energy}
\eeq
is the energy functional of the Dyson gas. In the large $N,M$ limit
(\ref{partition}) is governed by the saddle point of (\ref{energy}), 
namely, by a $\cdgc$ eigenvalue distribution $\{s_{\alpha}\}$ that 
satisfies
\beq
{\pa {\cal E}\over \pa s_{\alpha}} = V' (s_{\alpha}) -{r-1\over 
\sqrt{r}}~{1\over s_{\alpha}} - {2\over N\sqrt{r}}\sum_{\beta=1}^{~N~'} 
{1\over s_{\alpha}-s_{\beta}}\,.
\label{saddle}
\eeq
Here the prime over the sum symbol indicates that $\beta=\alpha$ is 
excluded from the sum.

We now turn our attention to the average eigenvalue density of $H$, 
which we 
may readily deduce\cite{BIPZ} from the averaged Green's function 
$G_{N,M}(z)$ 
in (\ref{gnm}). The $s_{\alpha}$ are eigenvalues of $\cdgc$. We thus 
calculate first $G_N(z^2)$, which according to (\ref{notation}), is 
given by 
\beq
G_N(w)={1\over N} \sum_{\alpha=1}^{N}\langle {1\over w-
s_{\alpha}}\rangle\,.
\label{gn}
\eeq
Here the angular brackets denote averaging with respect to 
(\ref{partition}).
By definition, $G_N(w)$ behaves asymptotically as 
\beq
G_N(w)\wasymptotic {1\over w}\,.
\label{wasymptotic}
\eeq
It is clear from (\ref{gn}) that for $s>0, \epsilon\rightarrow 0+$ we 
have 
\beq
G_N(s-i\epsilon)= {1\over N} {\rm P.P.}\sum_{\alpha=1}^N\langle {1\over 
s-s_{\alpha}}\rangle + {i\pi\over N}\sum_{\alpha=1}^N\langle\delta(s-
s_{\alpha})\rangle
\label{realim}
\eeq
where P.P. stands for the principal value. Therefore, the average 
eigenvalue 
density of $\cdgc$ is given by ${1\over \pi}~ {\rm Im}~ G_N(s-
i\epsilon)$. In the large $N,M$ limit, the real part of (\ref{realim}) 
is fixed by (\ref{saddle}), namely, 
\beq
{\rm Re}~ G_N(s-i\epsilon) = {1\over 2}\left[\sqrt{r}~ V'(s) - (r-
1)~{1\over s}\right]\,.
\label{real}
\eeq
The potential $V(s)$ in (\ref{pc}) clearly has at least one minimum for 
$s>0$, and will therefore cause the eigenvalues to coalesce into a 
single finite band or more along the real positive axis. Moreover, the 
${\rm log}~s$ term 
in (\ref{energy}) clearly implies that the $\{s_{\alpha}\}$ are repelled 
from the origin. We thus anticipate that the lowest band will be located 
at a finite distance from the origin $s=0$.

At this point we depart from discussing the general distribution and 
assume 
for simplicity that the probability distribution is given by the 
Gaussian distribution (\ref{complexhermitean}) with 
\beq
V(s) = m^2~s\,.
\eeq
In this case we expect the $\{s_{\alpha}\}$ to be contained in the 
single 
finite segment $0<b^2<s<a^2$, with $a>b>0$ yet to be 
determined.\footnote{We find below, of course, that $a$ and $b$ coincide 
with the expressions in (\ref{ab}).} This means that $G_N(w)$ should 
have a cut connecting $b^2$ and $a^2$. This conclusion, together with 
(\ref{real}) imply that $G_N(w)$ must be of the form
\beqast
G_N(w) = {1\over 2}\left[\sqrt{r}~ m^2 - (r-1)~{1\over w}\right] + 
F(w)~\sqrt{(w-b^2)(w-a^2)}\,,
\eeqast
where $F(w)$ is analytic in the $w$ plane (with the origin excluded.)
The asymptotic behavior (\ref{wasymptotic}) then fixes
\beq
F(w) = -{\sqrt{r}~m^2\over 2w}\quad\quad, \quad\quad a^2+b^2 = {2\over 
m^2}\left(\sqrt{r}+{1\over\sqrt{r}}\right)
\label{fab}
\eeq
and thus, 
\beq
G_N(w) = {1\over 2w} \left[ \sqrt{r}~m^2~w - r+1 - 
\sqrt{r}~m^2~\sqrt{(w-b^2)(w-a^2)}\right]\,.
\label{Gnw}
\eeq
The eigenvalue distribution of $\cdgc$ is therefore
\beq\label{rhotilde}
\tilde\rho (s) = {1\over \pi}~ {\rm Im}~ G_N(s-i\epsilon) =  
{\sqrt{r}~m^2\over 2\pi s} ~\sqrt{(s-b^2)(a^2 -s )}
\eeq
for $b^2<s<a^2$, and zero elsewhere.

We now substitute $G_N(z^2)$ from (\ref{Gnw}) into (\ref{simplergnm}) to 
obtain 
an expression for $G_{N,M}(z)$.  As we discussed in the introduction and 
in section $2$, $G_{N,M}(z)$ has a simple pole at $z=0$ with residue 
${M-N\over M+N} = {r-1\over r+1}$, which is the first term on the right 
side of (\ref{simplergnm}). We thus conclude from (\ref{simplergnm}) 
that $wG_N(w)$ must vanish at $w=0$, which in turn implies a second 
condition\footnote{Note that the Riemann sheet of the square root in 
(\ref{Gnw}) is such that $\sqrt{(0-b^2)(0-a^2)}=-ab$, as we already 
observed in section $2$.} on $a,b$, namely,
\beq
ab = {r-1 \over m^2~\sqrt{r}}\,.
\label{ab'}
\eeq
We are now able to fix $a$ and $b$ from (\ref{fab}) and (\ref{ab'}) and 
find
that they are given by (\ref{ab}). We thus find that $G_{N,M}(z)$ 
coincides 
with (\ref{gzhermitean}) and that the averaged eigenvalue density
of $H$ is the expression in (\ref{rhohermitean}).

We close this section by sketching the similar analysis of Gaussian 
random Hamiltonians made of real $M\times N$ blocks $C$. We parametrize 
the Gaussian real orthogonal ensemble by 
\beq\label{realsymmetric}
P(C)={1\over Z}{\rm exp}~[-{m^2\over 2}~\sqrt{NM}~\tr ~C^{T}C]
\eeq
with the partition function 
\beq
Z=\int \prod_{i=1}^M\prod_{\alpha = 1}^N d~C_{i\alpha}~{\rm exp}~[-
{m^2\over 2}~\sqrt{NM}~\tr~C^{T}C]\,.
\label{partition1real}
\eeq
The two point correlator associated with (\ref{realsymmetric}) is 
clearly
\beq\label{correlatorrealsymmetric}
\langle\cia\cjb\rangle\,=\,{1\over 
m^2\sqrt{MN}}~\delta_{ij}\delta_{\alpha\beta}\,.
\eeq
Note that (\ref{complexhermitean}) and (\ref{realsymmetric}) are 
conventionally parametrized in such a way that 
(\ref{correlatorhermitean}) and (\ref{correlatorrealsymmetric}) 
coincide.
 
The partition function for the corresponding Dyson gas reads 
\cite{periwal}
\beq
Z =\prod_{\alpha=1}^{N}~ \int\limits_{0}^{\infty}d s_{\alpha} ~{\rm 
exp}~[-{1\over 2}\sqrt{NM}~m^2~s_{\alpha}]  \prod_{\beta=1}^{N} 
s_{\beta}^{{M-N-1\over 2}} \prod_{1\leq\gam<\del\leq N} |s_{\gam}-
s_{\del}| \,.
\label{partitionreal}
\eeq
As before, the last two products constitute the Jacobian associated with 
polar coordinates. The energy functional ${\cal E}$ of the Dyson gas is 
now 
\beq
{\cal E} = {1\over 2}\left\{\sum_{\alpha = 1}^N \left( m^2~ s_{\alpha} - 
{r-1 -{1\over N}\over \sqrt{r}}~{\rm log} ~s_{\alpha}\right) - {1\over 
N\sqrt{r}}\sum_{1\leq\alpha<\beta\leq N} {\rm log}~ (s_{\alpha}-
s_{\beta})^2\right\}\,.
\label{energyreal}
\eeq
Thus, in the large $N,M$ limit, (\ref{energyreal}) becomes precisely one 
half
of the corresponding expression (\ref{energy}) for complex Hermitian 
matrices,
and our discussion following (\ref{saddle}) through (\ref{rhohermitean}) 
remains intact.
\pagebreak

\section{Kazakov's method extended to rectangular complex matrices}
\setcounter{equation}{0}

\vskip 5mm
{\bf 5.1 Contour integral}
\vskip 3mm
Gaussian matrix ensembles may be studied in many ways. Several years 
ago, Kazakov introduced a method \cite{Kazakov} for treating the usual 
Gaussian ensemble of random Hermitian
matrices, which was later extended and applied to a study of random 
Hermitian matrices made of square blocks\cite{bhznpb}. Here we 
generalize it to random Hermitian matrices made of rectangular blocks. 
It consists
of adding to the probability distribution a matrix source, which will
be set to zero at the end
of the calculation, leaving us with a simple integral
representation for
finite $N$.  As we will see,  one cannot let the source go to zero
before one reaches
the final step. We modify the probability distribution 
(\ref{complexhermitean})
of the matrix\footnote{For notational simplicity we set $m^2 = 1$ in 
(\ref{complexhermitean}) throughout this section} $\cdgc$ by adding a 
source $A$, an
$N\times N$ Hermitian matrix with eigenvalues
$(a_1,\cdots,a_N)$ :
\begin{equation}\label{AGaussian}
  P_A(C) = {1\over{Z_A}} {\exp} (-\sqrt{MN} {\rm Tr} C^{\dag}C -
\sqrt{MN} {\rm Tr} AC^{\dag}C). 
\end{equation}
Next we introduce the Fourier transform of the average
resolvent with this
modified distribution:
\begin{equation}\label{U}
U_A(t) =  \langle {1\over N} {\rm Tr} e^{itC^{\dag}C} \rangle_{_{A}}
\end{equation}
from which we recover the eigenvalue density 
 \begin{equation}\label{101}
  \tilde \rho(s) = \int_{-\infty}^{\infty} {dt\over{2\pi}} e^{-
its}U_0 (t) = \langle {1\over N} \tr \del\left(s-\cdgc\right)\rangle
\end{equation}
of $\cdgc$, after setting the source $A$ to zero.
Without loss of generality we can assume that $A$ is a
diagonal matrix. Let us now calculate $U_A(t)$. We first integrate over 
the $N\times N$ 
unitary matrix $U$ which diagonalizes $C^{\dag}C$ . This is
done through
the well-known Itzykson-Zuber integral over the unitary
group \cite{Itzykson}
\begin{equation}\label{IZ}
  \int dU \exp ({\rm Tr} AUBU^{\dag}) =  {\det [{\rm e} ^{a_{\al} 
b_{\bet}}]
\over
{\Delta(A) \Delta(B)}}
\end{equation}
where $\Delta(A)$ is the Vandermonde determinant
constructed with the
eigenvalues of $A$:
\begin{equation}\label{VdM}
  \Delta(A) = \prod_{\al <\bet} ( a_{\al} - a_{\bet})\,,
\end{equation}
$(b_1,\cdots,b_N)$ are the eigenvalues of $B$, and $\Delta(B)$ is the 
Vandermonde determinant built out of them. We are then led to
\begin{eqnarray}\label{111}
U_A(t) &=& {1\over {Z_A~\Delta(A)}}~{1\over N}
\sum_{\alpha=1}^{N}
\int ds_1\cdots ds_N~ e^{its_
{\alpha}}~\Delta(s_1,\cdots,s_N)\nonumber\\{}\nonumber\\
&\times & \left(\prod_{\bet = 1}^N s_{\bet}\right)^{M-N}~\exp
\left[-\sqrt{MN} \sum_{\gam=1}^N s_{\gam} \left(1 + 
a_{\gam}\right)\right]\,.
\end{eqnarray}
We now integrate over the $s_{\al}$'s. It is easy to 
prove (for example, by using the Faddeev-Popov method) that
\begin{eqnarray}\label{112}
&&\int ds_1\cdots ds_N ~\Delta(s_1,\cdots,s_N)~\left(\prod_{\bet = 1}^N 
s_{\bet}\right)^{M-N}~ \exp (-\sum_{\al = 1}^{N} s_{\al} b_{\al}) 
\nonumber\\
&&=  C_N ~{\Delta(b_1,\cdots,b_N)
\over (\prod_1^N ~b_{\al})^M}
\end{eqnarray}
where $C_N$ is a constant independent of the $b_{\alpha}$. Note that 
(\ref{112}) is valid also for $M=N$. With the normalization 
$U_{A}(0)=1$, we could always
divide, at any
intermediate step of the calculation, the expression we
obtain for
$U_{A}(t)$ by its value at $t=0$, and thus the overall
multiplicative
factors in (5.6) and (5.7) are not needed. 

We now apply this identity to the $N$ terms of (\ref{111}),
with
\begin{equation}\label{barelation}
b_{\beta}^{(\alpha)}(t) = \sqrt{MN} ( 1 + a_{\beta} - {it \over 
\sqrt{MN}}
\delta_{\alpha,\beta})
\end{equation}
and obtain
\begin{eqnarray}\label{113}
   U_A(t) &=&{1\over{N}}\sum_{\alpha=1}^{N}
\prod_{\beta=1}^{N} ({1 + a_{\beta}\over
{1 + a_{\beta} - {it\over\sqrt{MN}}\delta_{\alpha,\beta}}})^M
\prod_{\beta < \gamma}{a_\beta - a_\gamma
-{it\over\sqrt{MN}}(\delta_{\alpha,\beta}
-\delta_{\alpha,\gamma})\over{a_\beta -
a_\gamma}}\nonumber\\
&=& {1\over{N}}\sum_{\alpha=1}^N [ {1 + a_\alpha \over {1 +
a_\alpha - {it\over\sqrt{MN}}}}]^M \prod_{\gamma \neq \alpha}
( {a_\alpha  - a_\gamma - {it\over \sqrt{MN}}
\over {a_\alpha - a_\gamma}})
\end{eqnarray}
As a consistency check, note that for $M=N$, (\ref{113}) coincides with 
Eq. (3.9) of \cite{bhznpb}.
This sum over $N$ terms may be conveniently replaced by
a  contour integral
in the complex plane:
\begin{equation}\label{114}
U_A(t) =   {i~\sqrt{M\over N}\over{t}}\oint {du\over{2\pi i}}\left({1 + 
u
\over{1 + u -
{it\over\sqrt{MN}}}}\right)^M \prod_{\gamma=1}^N {u - a_\gamma
- {it\over
\sqrt{MN}}\over{ u -a_\gamma}}
\end{equation}
in which the contour encloses all the $a_\gamma$'s and no
other singularity.
  It is  now, and only now, possible to let all the
$a_\gamma$'s go to
zero. We thus obtain a simple expression for
$U_0(t)$,
\begin{equation}\label{115}
   U_0(t) = {i~\sqrt{{M\over N}}\over t }  \oint {du\over 2\pi 
i}{\left(1 -
{it\over u\sqrt{MN}}\right)^N\over
\left(1 - { it\over (1+u)\sqrt{MN}}\right)^M}\,.
\end{equation}
Note that this representation of $U_0(t)$ as a contour integral  over
one single
complex variable is  exact for any
finite $M, N$, including $M=N=1$.
\vskip 5mm
{\bf 5.2 The density of states} 
\vskip 3mm
In the large $M, N$ limit (with finite $r = {M\over N}$), for finite 
$t$, the integrand in (\ref{115}) 
becomes $e^{it\left({\sqrt{r}\over u+1} - {1\over u\sqrt{r}}\right)}$
and therefore $U_0(t)$ approaches
\begin{equation}\label{116}
      U_0(t) = {\sqrt{r}\over it}~\oint {du\over{2\pi i}}
e^{it\left[{\sqrt{r}\over\left(1-u\right) } + {1\over u\sqrt{r}}\right]}
\end{equation}
where we changed $u$ into $-u$. 

Setting $ z = {1\over u\sqrt{r} } + {\sqrt{r}\over 1 - u}$ we change 
variables 
to 
\begin{equation}\label{117}
     u = {z -\sqrt{r} + {1\over\sqrt{r}} - \sqrt{\left(z -\sqrt{r} + 
{1\over\sqrt{r}}\right)^2 - {4z\over\sqrt{r}}} \over 2 z}\end{equation}
Then the integral of (\ref{116}) becomes, after an
integration by parts,
\begin{eqnarray}\label{118}
     U_0(t) &=& {\sqrt{r}\over it}~ \oint {dz\over{2\pi i}} ~{du\over 
dz}~ e^{itz}   = -\sqrt{r}~ \oint {dz\over{2\pi i}}~ 
u(z)~e^{itz}\nonumber\\
          &=& - \sqrt{r} \oint {dz\over{2\pi i}}~ \left[z -\sqrt{r} + 
{1\over\sqrt{r}} - \sqrt{\left(z -\sqrt{r} + {1\over\sqrt{r}}\right)^2 - 
{4z\over\sqrt{r}}}~\right]~{e^{itz}  \over 2 z} ~
\nonumber\\
   &=& {\sqrt{r}\over 2\pi} \int_{b^2}^{a^2} ~{dx\over x}~ 
\sqrt{\left(a^2-x\right)\left(x-b^2\right)}~ e^{itx}
\end{eqnarray}
where $a$ and $b$ are given in (\ref{ab}). Therefore, we have from 
(\ref{101})
\begin{eqnarray}\label{119}
    \tilde \rho (s) &=& \int {dt\over{2\pi}}
       e^{-its} U_0(t)\nonumber\\
     &=& {\sqrt{r}\over 2\pi} {\sqrt{\left(a^2-s\right)\left(s-
b^2\right)}\over s}
\end{eqnarray} 
for $b^2\leq s\leq a^2$, and zero elsewhere. This expression coincides 
with (\ref{rhotilde}) as expected. 
We observe from (\ref{simplergnm}) that $\rho(\lambda)$ 
and $\tilde\rho(s)\equiv\tilde\rho(\lambda^2)$ are related by
\beq\label{rhorhotilde}
\rho(\lambda) = {r-1\over r+1}\delta(\lambda) + {2|\lambda|\over 
r+1}\tilde\rho (\lambda^2)\,.
\eeq
Substituting  (\ref{119}) into (\ref{rhorhotilde}) we obtain 
(\ref{rhohermitean}) once again, as we should.

\vskip 5mm
{\bf 5.3 The edges of the eigenvalue distribution}
\vskip 3mm
 It is easy to apply this same method for studying the
cross-over at the edges of the eigenvalue distributions 
(\ref{rhohermitean}) or (\ref{119}), namely, in the vicinity of the end 
points $s=a^2$ and $s=b^2$.
To this end, we observe from (\ref{101}) and (\ref{115}) that 
\beq\label{drhods}
{\pa \tilde\rho(s)\over \pa s} = \sqrt{r}~\int_{-\infty}^{\infty} 
{dt\over{2\pi}} e^{-its}~\oint {du\over 2\pi i}{\left(1 -
{it\over N u\sqrt{r}}\right)^N\over
\left(1 - { it \sqrt{r}\over M (1+u)}\right)^M}
\end{equation}
where the purpose of the $s$ derivative is to get rid of the simple pole 
at $t=0$ in (\ref{115}). By changing $t$ to $\sqrt{MN}~t$ and then $t$ 
to $t+iu$, as well as $u$ to $-iu$, we obtain the factorized expression 
\beq\label{drhodsfactorised}
{\pa \tilde\rho(s)\over \pa s} = -iM~\left(\int_{-\infty}^{\infty} 
{dt\over{2\pi}} e^{-i\sqrt{MN}~ts}~{t^N\over 
\left(t+i\right)^M}\right)\cdot\left(
\oint {du\over 2\pi i}~e^{i\sqrt{MN}~us}~{\left(u+i\right)^M \over 
u^N}\right)\end{equation}
The advantage of (\ref{drhodsfactorised}) is that it is relatively easy 
to 
study its large $N,M$ behavior by saddle point techniques. We observe 
that the
$t$ integral may be written as 
\beq\label{inm}
I_{N,M} = \int_{-\infty}^{+\infty} dt e^{-\sqrt{NM}~S_{eff}}
\eeq
where $S_{eff}$ is given by
\begin{equation}\label{seff}
    S_{eff} = i~s~t + \sqrt{r}~{\rm log}~ (t+i) - {1\over \sqrt{r}}~{\rm 
log}~ t\,. \end{equation}  
Similarly, the integrand of the $u$ integration is 
$e^{\sqrt{NM}~S_{eff}}$.
Thus, the large $N,M$ behavior of (\ref{drhods}) is determined by the 
saddle points of a single function $S_{eff}$. Consider the $t$ integral 
(\ref{inm})
first. It has two saddle points $t_c$ at 
\begin{equation}\label{tc}
   t_c = {s - ab \pm \sqrt{\left(s-a^2\right)\left(s-b^2\right)}\over 
2is}
\end{equation}
where $a, b$ are given in (\ref{ab}). The interesting situation occurs 
when these two saddle points become degenerate, namely at the endpoints 
$s=\lambda^2 = a^2$ and $s=\lambda^2=b^2$. We thus investigate 
(\ref{drhods})
at the vicinity of these points, by focusing on these regions. 
Let us consider the neighborhood of $\lambda=a$ first (the cross over 
behavior around $\lambda=-a$ is simply the mirror image thereof.) We 
introduce the 
scaled variables 
\begin{eqnarray}
    \lambda &=&  a + N^{-\alpha} x,\nonumber\\
    t &=& - i{1\over \sqrt{r}+1} + N^{-\beta} \tau\,,
\end{eqnarray}
with $\alpha, \beta$ to be determined, and expand $S_{eff}$ up to 
$\tau^3$. This  leads to
\begin{eqnarray}\label{taus}
S_{eff}(t) &=& S_{*} + 2~r^{-{1\over 4}}~N^{-\alpha} x
\nonumber\\
&+&
{i\over 3}~ a^4~\tau^3 ~N^{-3\beta}
   + 2~i~a~ N^{-\alpha - \beta} \tau x + \cdots
\end{eqnarray}
where $S_{*}$ is the value of $S_{eff}$ at the critical point, and the 
ellipsis stand for terms of $\cO (N^{-2\alpha})$. We thus find that 
there is a large 
$N$, finite $x$ limit,
provided we
fix the two unknown exponents $\alpha$ and $\beta$ to
\begin{equation}
   \alpha = {2\over{3}}, {\hskip 5mm} \beta = {1\over{3}}
\end{equation}
We repeat this for the $u$-integral of (\ref{drhodsfactorised}).  We
then find that
the leading terms of (\ref{taus}) of order 1, as well as the
term   $2x
N^{-2/3}$,  cancel  with terms of opposite
signs  in the $u$-integral.
Thus we obtain the following equation for the density of
state near the
critical value $s = a^2$ or $\lambda = \pm a$,  
\begin{equation}\label{edge Airy}
{\partial \tilde \rho(\lambda^2)\over{\partial \lambda^2}}
= - M^{1\over{3}} \left({\sqrt{r}\over a^4}\right)^{2\over 3}~
\Big|Ai\left[2\left({r\over a}\right)^{{1\over 3}}N^{2\over{3}}(\lambda 
\mp a)\right]\Big|^2
\end{equation}
where the Airy function $Ai(z)$ is defined as
\begin{equation}
     Ai[ (3\alpha)^{-1/3} x] = {(3\alpha)^{1/3}\over{\pi}}
\int_0^\infty \cos (\alpha t^3 + xt)dt.
\end{equation}
The Airy function in (\ref{edge Airy}) is  smoothly decreasing for 
$|\lambda| > a$ but
it oscillates for $|\lambda| < a$.

Investigation of the behavior of (\ref{drhodsfactorised}) near the other 
critical points $\lambda=\pm b$ proceeds similarly. Concentrating on 
$\lambda=b$ we introduce the scaling variables 
\begin{eqnarray}\label{bpoint}
    \lambda &=&  b + N^{-\alpha} x,\nonumber\\
    t &=& {i\over \sqrt{r}-1} + N^{-\beta} \tau
\end{eqnarray}
and find that  there is a large 
$N$, finite $x$ limit, provided we fix the two unknown exponents 
$\alpha$ and $\beta$ to the same values as before. Thus, the crossover 
behavior of the density of states arround $\lambda=\pm b$ is governed by 
the Airy function as well, for any $r>1$.

A new phenomenon appears, however, if we also take the limit 
$r\rightarrow 1$. It is easy to see, by rescaling $\tau$ in the 
expansion of $S_{eff}$ into  
\beq
T=(\sqrt{r} - 1)\tau\,,
\eeq
that the Airy function behavior of ${\pa \tilde\rho (\lambda^2)\over \pa 
\lambda^2}$ near $\lambda=\pm b$ breaks down as $r\rightarrow 1$. 
Indeed, from previous work \cite{Verbaarschot, amb, ambjorn, NS, Nforr, 
AST, bhznpb, Nishigaki} we know that the oscillations near the origin in 
the density of the eigenvalues of matrices built out of square blocks 
$(r=1)$ are governed by the Bessel function and not by the Airy function.

\vskip 10mm
\begin{center}
{\bf ACKNOWLEDGEMENTS}
\end{center}

This work  is supported in part by the National Science Foundation under 
Grant No. PHY89-04035.


\newpage
\setcounter{equation}{0}
\renewcommand{\theequation}{A.\arabic{equation}}
{\bf Appendix : {The Central Limit Theorem - A Renormalization Group 
Proof}}
\vskip 5mm

As a simple, but perhaps amusing exercise we use the large $N$ 
renormalization group discussed in Section 3 to prove the celebrated 
central limit theorem
of Gauss. 

Consider a set of $N$ independent random variables $\{x_1, x_2, \cdots, 
x_N\}$ which are distributed according to some distribution function 
\beq\label{A0}
Q_N(x_1, \cdots, x_N) = \prod_{i=1}^N Q(x_i)\,.
\eeq
In order to be consistent with our normalization conventions in Section 
3, we normalize
this distribution function such that 
\beq
\langle x_i\rangle = 0\,,\quad\quad \langle x_i x_j\rangle = {\si^2\over 
N^{2\beta}}~\delta_{ij}
\label{A1}
\eeq
where $\beta >0$ is yet to be determined. Thus, a typical term drawn 
from 
$Q_N(x)$ is of the order $\si~N^{-\beta}$.
We wish to calculate the distribution function of the sum of these 
random numbers, namely, the quantity 
\beq
P_N (s,\si) =\Big \langle\delta\left(s-\sum_{i=1}^N 
x_i\right)\Big\rangle_{N}
\label{A2}
\eeq
where $\langle\cdot\rangle_{N}$ denotes averaging with respect to 
$Q_N(x)$. In principle, $P_N$ depends upon all the cumulants of 
$Q_N(x)$, but we expect
that the large $N$ limit of $P_N$ will depend only upon $\si$. Following 
our discussion in Section 3, we now consider a set of $N+1$ random 
variables whose distribution function $Q_{N+1}(x)$ is normalized such 
that
\beq
\langle x_i\rangle = 0\,,\quad\quad \langle x_i x_j\rangle = {\si^2\over 
(N+1)^{2\beta}}~\delta_{ij}\,.
\label{A3}\eeq
Then,
\beqra
P_{N+1} (s,\si) &=&  \Big\langle\delta\left(s-\sum_{i=1}^{N} x_i - 
x_{N+1}\right)\Big\rangle_{N+1}\nonumber\\{}\nonumber\\ &=& 
\Big\langle\delta\left(s-\sum_{i=1}^{N} 
x_i\right)\Big\rangle_{N+1}\nonumber\\{}\nonumber\\ &+& ~{\si^2\over 
2\left(N+1\right)^{2\beta}}~{\pa^2\over \pa 
s^2}~\Big\langle\delta\left(s-\sum_{i=1}^{N} x_i\right)\Big\rangle_{N+1} 
+\cdots
\label{A4}
\eeqra
where we used (\ref{A3}). The ellipsis stand for cumulants 
of order higher than two, which are clearly suppressed by powers of 
$N^{-\beta}$, and we neglect them henceforth.  Comparing (\ref{A1}) and 
(\ref{A3}) we also see that 
\beq
\Big\langle\delta\left(s-\sum_{i=1}^{N} x_i\right)\Big\rangle_{N+1} = 
P_N(s,\si')\label{A5}
\eeq
with 
\beq
\si' = \left(N\over N+1\right)^{\beta}~\si = (1-{\beta\over N})~\si 
+\cdots
\label{A6}
\eeq
We now use (\ref{A5}) and (\ref{A6}) to rewrite (\ref{A4}) as
\beq
P_{N+1} (s,\si) = \left[ 1 - {\beta\over N}~\si~{\pa\over \pa \si} + 
{\si^2\over 2 N^{2\beta} }~{\pa^2\over \pa s^2}\right] P_N 
(s,\si)\label{A7}
\eeq
where we neglected terms of $\cO ({1\over N^{2\beta +1}})$. We observe 
from 
(\ref{A7}) that variations of $\si$ are as important as variations of 
$s$ in 
the large $N$ limit only if 
\beq
\beta = {1\over 2}
\label{A8}
\eeq
which fixes $\beta$. We thus conclude that 
\beq
N~{\pa P_N\over N} = {\si\over 2} \left[\si~{\pa^2\over \pa s^2}  - 
{\pa\over \pa \si}\right] P_N (s,\si)\,.\label{A9}\eeq
The left hand side of (\ref{A8}) must vanish if $P_N$ has a large $N$ 
limit
\beq
P_N(s,\si)\nasymptotic P(s,\si)\,,
\label{A10}
\eeq
and thus
\beq\label{A11}
\left[\si~{\pa^2\over \pa s^2}  - {\pa\over \pa \si}\right] P (s,\si) = 
0\,.
\eeq
A simple scaling argument, similar to the one invoked in Section 3, 
leads to
the relation  
\beq
P(s,\si) = {1\over \si} P({s\over\si},1)
\label{A12}\eeq
which implies that 
\beq\label{A13}
\si~{\pa\over \pa \si} ~P  = - P - s~{\pa \over \pa s}~P\,.\eeq
Substituting (\ref{A13}) in (\ref{A11}) we finaly obtain the 
differential equation
\beq\label{A14}
\left(\si^2~{\pa^2\over \pa s^2} + s {\pa\over \pa s} +1\right)P(s,\si) 
= 0\,.
\eeq
We solve (\ref{A14}) and find that its normalized solution is the 
Gaussian
distribution
\beq\label{A15}
P(s,\si) = {1\over \sqrt{2\pi}~\si}~{\rm exp} \left(-{s^2\over 
2\si^2}\right)
\eeq
which is the statement of the central limit theorem. The proof of the 
central limit theorem presented here is not any simpler than the 
conventional proof found in textbooks.

The generalization of this proof to the case\cite{blue} of adding a 
large 
number $N$
of $K\times K$ matrices $\{\phi_1, \cdots, \phi_N\}$ is straightforward. In 
this case $s$ and $P(s,\si)$ are $K\times K$ matrices. We 
take 
these matrices to be real (the Hermitian case can be treated similarly.) 
Then (\ref{A14}) becomes
\beq\label{A16}
\left(\si^2~{\pa^2\over \pa s^{\mu}_{\nu} \pa s^{\nu}_{\mu}} + 
s^{\mu}_{\nu} {\pa\over \pa s^{\mu}_{\nu}} +1\right)P(s,\si) = 0
\eeq
where $\mu, \nu$ are indices of the $K\times K$ matrices (repeated 
indices are summed over.) The normalized solution of (\ref{A16}) is the 
Gaussian
distribution
\beq\label{A17}
P(s,\si) = \left(\sqrt{2\pi}~K~\si\right)^{-K^2}~{\rm exp} \left(- {\tr~ 
s^2\over 2K^2\si^2}\right)\,.
\eeq


\newpage
\begin{thebibliography}{99}
\bibitem{Verbaarschot}J.~J.~M.~Verbaarschot, {\it Nucl.
~Phys.} {\bf B426}
(1994)
559.
\bibitem{amb}  J.~Ambj{\o}rn, ``Quantization of Geometry,"
in Les Houches
1990, edited by J.~Dalibard et al. See section 4.3 and
references
therein.
\bibitem{ambjorn}
J.~Ambj{\o}rn, J.~Jurkiewicz, and Yu.~M.~Makeenko, {\it
Phys.~
Lett.} {\bf B251} (1990) 517.
\bibitem{NS}  K.~Slevin and T.~Nagao, {\it Phys.~Rev.~Lett.}
{\bf 70}
(1993) 635, {\it
Phys.~Rev.~}
{\bf B 50 } (1994) 2380.
T.~Nagao and K.~Slevin, {\it J. Math.
          Phys.} {\bf 34} (1993) 2075, 2317.


\bibitem{Nforr} T. ~Nagao and P. ~J. ~Forrester, {\it
Nucl.~Phys.} {\bf B 435} (FS)
(1995)
401.

\bibitem{AST}  A.~V.~Andreev, B.~D.~Simons and
N.~Taniguchi,
{\it Nucl.~Phys.} {\bf B 432} (1994) 485.
\bibitem{bhznpb} E.~Br\'ezin, S.~Hikami and A.~Zee, {\it
Nucl. ~Phys.}{\bf B 464} (1996) 411.
\bibitem{Nishigaki} S. ~Nishigaki  preprint, hep-th/9606099.
\bibitem{VerbaarschotZ} J.~J.~M.~Verbaarschot and
I.~Zahed, {\it Phys.~Rev.~Lett.} {\bf 70} (1993) 3852.
 
\bibitem{Zahed}  J.~Jurkiewicz, M.A.~Novak, and  I.~Zahed, preprint hep-
ph/9603308.   M.A.~Novak, G.~Papp and  I.~Zahed, preprint hep-
ph/9603348.
\bibitem{HZ} S. Hikami and A. Zee, {\it Nucl. ~Phys.} {\bf B 446
}, (1995)
337.

\bibitem{HSW} S.~Hikami, M.~Shirai and F.~Wegner,
{Nucl.~Phys.} {\bf B
408} (1993) 415.
\bibitem{hanna} C.B.~Hanna, D.P. ~Arovas, K. ~Mullen and S.M.~Girvin,
cond-mat 9412102.

\bibitem{bzw} E.~Br\'ezin and A. Zee, {\it Phys. ~Rev.} {\bf E 49
}, (1994)
2588.
\bibitem{french} E.~Br\'ezin and A. Zee, {\it Comp. ~Rend. ~Acad. 
~Sci.}, (Paris)  {\bf 317}, (1993)
735.
\bibitem{rg} E. ~Br\'ezin and J. ~Zinn-Justin, {\it Phys. Lett B} {\bf 
288}, 54 (1992).\\
S. ~Higuchi, C.~Itoh, S.~Nishigaki and N.~Sakai,  ~~ {\it Phys.~Lett.} 
{\bf B 318}, (1993) 63; ~~ {\it Nucl. ~Phys.} {\bf B 434}, (1995) 283, 
Err.-ibid. {\bf B 441}, (1995) 405.
\bibitem{daz} J.~D'Anna and A. ~Zee, {\it Phys. ~Rev.} {\bf E 53}, 
(1996)
1399.
 
\bibitem{periwal} A.~ Anderson, R.~C. Myers and V.~ Periwal, {\it Phys. 
Lett.}{\bf B 254} (1991) 89, ~~ {\it Nucl. ~Phys.} {\bf B 360}, (1991) 
463.\\
R.~C. Myers and V.~ Periwal, {\it Nucl. ~Phys.} {\bf B 390}, (1991) 716.
\bibitem{Kazakov} V.~A.~Kazakov, {\it Nucl.~Phys.} {\bf B
354}
(1991) 614.
\bibitem{wigner} E.P.~ Wigner, Can. Math. Congr. Proc. p.174, University 
of Toronto Press (1957), reprinted in 
C. E.~ Porter, Statistical Theories of Spectra:
Fluctuation (Academic, New york, 1965). 
See also  M.~L.~Mehta, Random Matrices (Academic,
New York, 1991).
\bibitem{BIPZ} E.Br\'ezin, C. Itzykson, G. Parisi and J.~-B.
Zuber, {\it Comm.~Math.~Phys.} {\bf 59}, 35 (1978).
\bibitem{Itzykson} C.~Itzykson and J.~-B.~Zuber, {\it
J.~Math.~Phys.}
{\bf 21} (1980) 411.
\bibitem{blue} A.~Zee, {\it Nucl.~Phys.} {\bf B
354}, in press.

\end{thebibliography}
\end{document}